\documentclass[aps,prl,twocolumn,superscriptaddress,notitlepage]{revtex4-1}
\usepackage[utf8]{inputenc}
\usepackage[english]{babel}
\usepackage{amsmath}
\usepackage{amsfonts}
\usepackage{amssymb}
\usepackage{graphicx}
\usepackage{hyperref}
\usepackage{pgf}
\usepackage{braket}
\usepackage{color,soul}

\DeclareUnicodeCharacter{2212}{$-$}

\newcommand{\Ic}{\ensuremath{I_\text{c}}}
\newcommand{\Vg}{\ensuremath{V_\text{G}}}
\newcommand{\fq}{\ensuremath{f_\text{Q}}}
\newcommand{\Ej}{\ensuremath{E_\text{J}}}
\newcommand{\Ec}{\ensuremath{E_\text{C}}}
\newcommand{\EjEc}{\ensuremath{E_\text{J}/E_\text{C}}}
\newcommand{\Qthr}{QA}
\newcommand{\Qone}{QB}

\let\pgfimageWithoutPath\pgfimage 
\renewcommand{\pgfimage}[2][]{\pgfimageWithoutPath[#1]{Figures/#2}}

\begin{document}

\title{Realization of a Carbon-Nanotube-Based Superconducting Qubit}

\author{Matthias Mergenthaler}
\email{mme@zurich.ibm.com}
\affiliation{Clarendon Laboratory, Department of Physics, University of Oxford, Oxford OX1 3PU, United Kingdom}
\affiliation{Department of Materials, University of Oxford, Oxford OX1 3PH, United Kingdom}
\affiliation{IBM Research Zurich, S\"{a}umerstrasse 4, 8803 R\"{u}schlikon, Switzerland}

\author{Ani Nersisyan}
\affiliation{Clarendon Laboratory, Department of Physics, University of Oxford, Oxford OX1 3PU, United Kingdom}
\affiliation{Rigetti Computing, 2919 Seventh Street, Berkeley, CA 94710}

\author{Andrew Patterson}
\affiliation{Clarendon Laboratory, Department of Physics, University of Oxford, Oxford OX1 3PU, United Kingdom}

\author{Martina Esposito}
\affiliation{Clarendon Laboratory, Department of Physics, University of Oxford, Oxford OX1 3PU, United Kingdom}

\author{Andreas Baumgartner}
\affiliation{Department of Physics, University of Basel, Klingelbergstrasse 82, CH-4056 Basel, Switzerland}

\author{Christian Schönenberger}
\affiliation{Department of Physics, University of Basel, Klingelbergstrasse 82, CH-4056 Basel, Switzerland}

\author{G. Andrew D. Briggs}
\affiliation{Department of Materials, University of Oxford, Oxford OX1 3PH, United Kingdom}

\author{Edward A. Laird}
\affiliation{Department of Physics, Lancaster University, Lancaster LA1 4YB, United Kingdom}
\affiliation{Department of Materials, University of Oxford, Oxford OX1 3PH, United Kingdom}

\author{Peter J. Leek}
\email{peter.leek@physics.ox.ac.uk}
\affiliation{Clarendon Laboratory, Department of Physics, University of Oxford, Oxford OX1 3PU, United Kingdom}
	
\date{\today}

\begin{abstract}
Hybrid circuit quantum electrodynamics (QED) involves the study of coherent quantum physics in solid state systems via their interactions with superconducting microwave circuits. Here we present an implementation of a hybrid superconducting qubit that employs a carbon nanotube as a Josephson junction. We realize the junction by contacting a carbon nanotube with a superconducting Pd/Al bi-layer, and implement voltage tunability of the qubit frequency using a local electrostatic gate. We demonstrate strong dispersive coupling to a coplanar waveguide resonator via observation of a resonator frequency shift dependent on applied gate voltage. We extract qubit parameters from spectroscopy using dispersive readout and find qubit relaxation and coherence times in the range of $10-200~\rm{ns}$.
\end{abstract}
\maketitle

Circuit quantum electrodynamics (cQED) with superconducting circuits~\cite{Wallraff2004} has been a remarkably successful platform for on-chip quantum optics and quantum information processing research~\cite{Wendin2017}. Hybrid superconducting circuits provide the platform to access coherent quantum properties of other systems via their interactions with microwave photons or artificial atoms~\cite{Xiang2013,Kurizki2015,Cottet2017}. In recent years, a variety of hybrid superconducting qubits have been realised by replacing the conventional aluminum (SIS) Josephson junctions (JJ) with normal- or semiconductor-based (SNS) JJs, such as InAs nanowires~\cite{DeLange2015,Larsen2015a}, InGaAs heterostructures~\cite{Casparis2018a} and graphene~\cite{Kroll2018a,Wang2018}. For these SNS JJs the normal- or semiconductor is contacted with a superconducting material enabling a supercurrent to flow due to the superconducting proximity effect~\cite{Meissner1960}. Cooper pair transport in such devices is described by Andreev reflections~\cite{Andreev1965,DeGennes1963,Saint-James1964}. The conductance of semiconductors can be adjusted by applying a voltage to a nearby gate-electrode, which for an SNS JJ tunes the Cooper-pair transport and hence the Josephson energy of the junction.

A strong technical motivation for these new semiconductor-superconductor hybrid JJ qubits is to realize gate voltage tunable qubits and hence eliminating decoherence due to magnetic flux noise. Further, electric fields are much easier to localise compared to magnetic fields, which makes complex multi-qubit devices simpler to engineer. Additionally, qubit operation in moderate magnetic fields, for example to explore interactions with different spin systems, can be made possible due to the robustness of these hybrid JJs to magnetic field~\cite{Luthi2018}. 

A variety of novel physical investigations could be envisaged with carbon nanotube (CNT)-based superconducting qubits. Using a CNT as the junction allows to make use of its exceptional mechanical properties, which could offer a potential platform for creating quantum interference between a qubit and mechanical motion~\cite{Khosla2018}. Further, ultra-clean CNTs offer ballistic transport characteristics and could be combined with growth from a nuclear spin free C$_{12}$ precursor. This would allow for perfect, defect free JJs as opposed to conventional Al JJs with an amorphous tunnel barrier. Further, this could have a positive impact on qubit coherence via elimination of two-level fluctuator defects in the amorphous oxide~\cite{Shnirman2005,Oh2006,Muller2017}. Furthermore, hybrid devices incorporating proximitized CNTs would allow to study Andreev levels~\cite{Gramich2017,Tosi2019,Janvier2015} and they are also predicted to carry Majorana fermions~\cite{Klinovaja2012,Sau2013,Marganska2018}, which could be beneficial for topological quantum computing~\cite{Kitaev2003}.
\begin{figure}[h!t]
	\centering
	\includegraphics[width=\columnwidth]{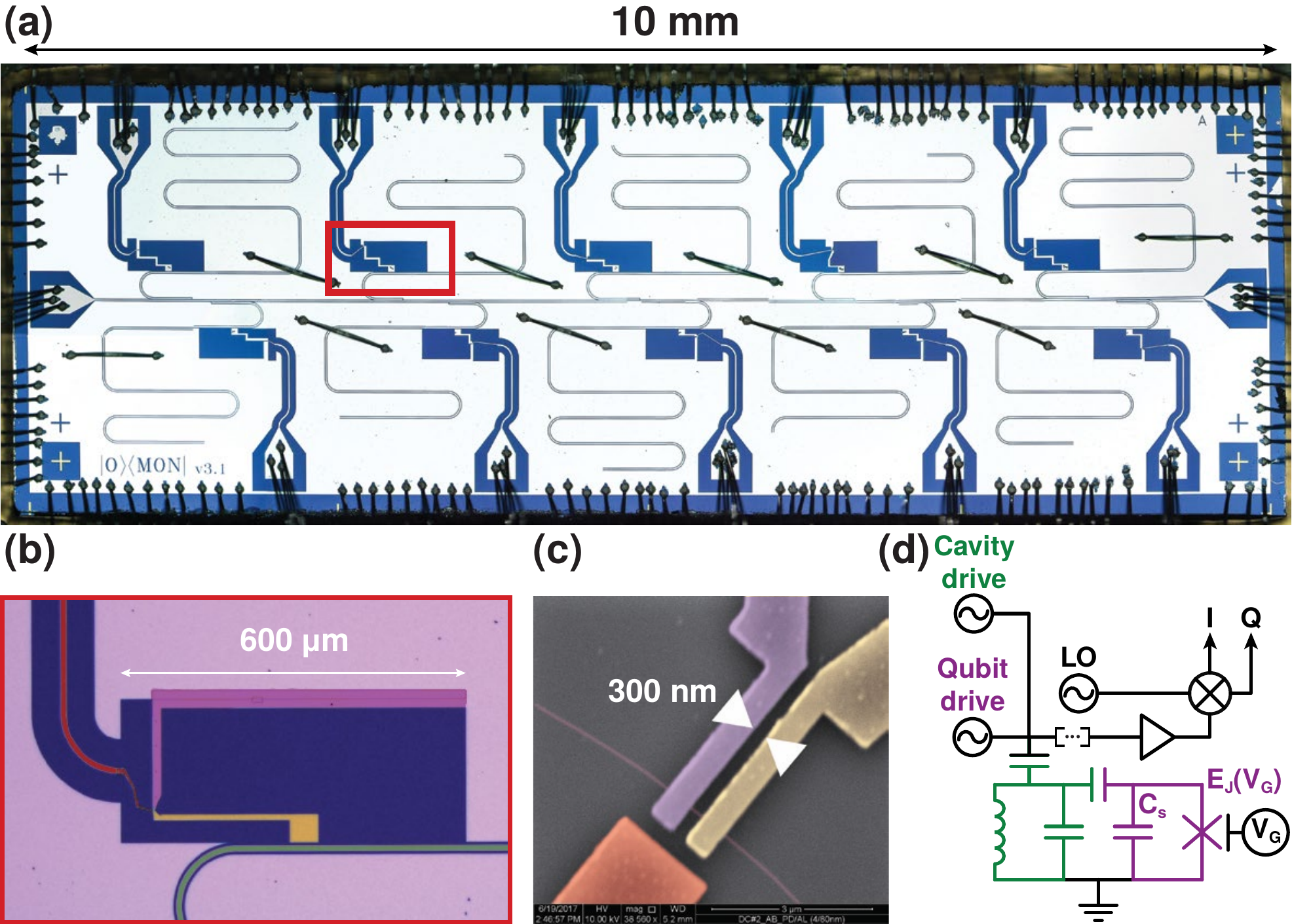}
	\caption{Carbon nanotube superconducting qubit device. (a) Optical image of the qubit device chip. A coplanar microwave transmission line in the center addresses 10 multiplexed $\lambda/4$ resonators with different resonance frequencies. Each resonator has a cut-out in the ground plane close to its electric field anti-node for qubit fabrication and a single DC line allowing voltage tunability of the qubit frequency. (b) False color optical image of a single qubit with the island (yellow) coupled to the resonator (green) and the other side shorted to ground (purple). A side gate (red) is used to tune the qubit frequency. (c) False color SEM image of a CNT (pink) contacted with two superconducting contacts (yellow/purple) separated by $300$~nm and a side gate (red). (d) Electrical circuit diagram of the device along with a sketch of its readout and control circuitry. The qubit (purple) is capacitively coupled to the resonator (green). A side gate with applied voltage \Vg~tunes the Josephson energy \Ej~of the qubit. The resonator itself is capacitively coupled to a transmission line which is used to send microwave tones to the qubit and its response is measured using a standard heterodyne detection scheme.}
	\label{fig1}
\end{figure}

In the work presented here proximitized CNTs are used as the JJ in a common planar 2D superconducting qubit architecture and their performance as a qubit is analyzed via a coupled microwave resonator. Resonator and qubit spectroscopy are performed as a function of applied gate voltage and strong dispersive coupling on the order of 100~MHz is observed. Power dependent qubit spectroscopy is used to extract the Josephson energy \Ej~and charging energy \Ec~of the qubits. Further, the coherence is investigated and $T_1$ and $T_2$ times in the range of 10-200~ns are observed.  

Fig.~\ref{fig1} shows images and a circuit diagram of the device studied. The chip consists of 10 $\lambda/4$ resonators at different frequencies, multiplexed through a single coplanar microwave transmission line (Fig.~\ref{fig1}~(a)). Close to each resonator's electric field anti-node, qubits are fabricated and a dedicated DC electrostatic gate is used for control of the chemical potential of the CNT (Fig.~\ref{fig1}~(b-c)). 
  
The system is cooled below 20~mK in a dilution refrigerator and measured using standard cQED measurement techniques, see Fig.~\ref{fig1}~(d). First, the transmission spectrum of a device is measured via the feedline to identify the individual resonance frequencies of the 10 resonators, at which a narrow ($\sim 1$~MHz) absorption dip is observed. Subsequently, $S_{21}$ spectroscopy as a function of gate voltage $\Vg$ of each individual resonance is performed and resonators which exhibit a clear gate-dependent resonance frequency are selected for further investigations as these potentially correspond to working CNT-qubits. Usually 20-50\% of all devices on one chip show this dependence. From these, two devices showing similar gate-dependent behaviour are carefully characterized, hereafter labeled as device QA and QB. 

\begin{figure}[t!]
	\centering
	\includegraphics{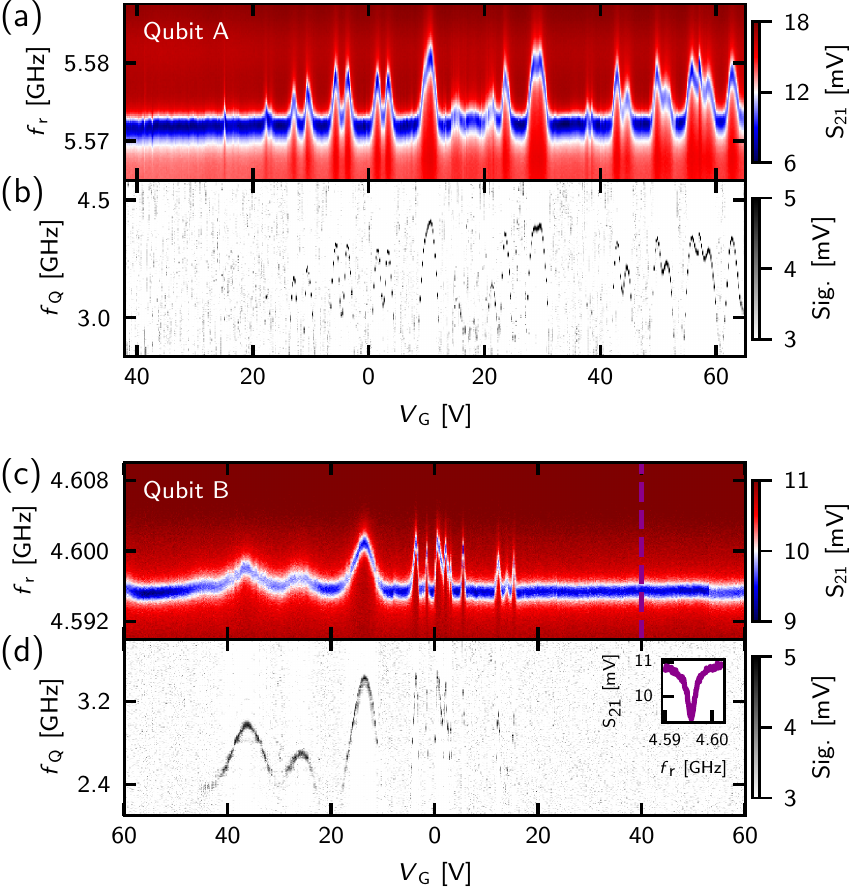}
	\caption{Resonator and qubit spectroscopy as function of applied gate voltage $\Vg$ for devices QA ((a) and (b)) and QB ((c) and (d)). (a) Resonator spectroscopy of device QA as a function of $\Vg$. (b) Qubit spectroscopy of device QA as a function of $\Vg$. (c) Resonator spectroscopy of device QB as a function of $\Vg$. (d) Qubit spectroscopy of device QB as a function of $\Vg$. Inset: Line-cut through spectrum indicated in (c).}
	\label{fig2}
\end{figure}

Fig.~\ref{fig2}~(a) shows resonator spectroscopy as a function of DC voltage applied to the gate electrode ($V_\textrm{G}$) on device QA. At a single gate-voltage \Vg, an absorption line corresponding to the resonator is observed (inset Fig.~\ref{fig2}~(d)). Tuning \Vg~the absorption line moves in frequency, exhibiting a broad gate region ($\Vg<-30$~V) with an approximately constant resonator frequency of $f_0=5.572$~GHz (bare resonator) and quasi-periodic excursion to higher frequencies for $\Vg >-20$~V. In cQED this is indicative of the resonator being dispersively coupled to a qubit with tunable transition frequency $f_\textrm{Q}<f_0$. Here, the resonator frequency is observed to vary by $\sim 10$~MHz, which is consistent with the dispersive regime of cQED~\cite{Blais2004}.

In a second measurement, we carry out spectroscopy to identify the potential qubit's frequency. Here the cavity drive is tracking the particular resonance frequency $f_r$ at each $V_\textrm{G}$ and simultaneously a spectroscopic probe tone is fed onto the input line (qubit drive). While the qubit drive is swept in frequency the amplitude response at $f_r$ is measured. This measurement performed on device QA is presented in Fig.~\ref{fig2}~(b). A spectroscopic response is observed for $V_\textrm{G}>-30$~V. The values of $V_\textrm{G}$ exhibiting a spectroscopic response coincide exactly with the values exhibiting shifts in $f_r$, cf. Fig.~\ref{fig2}~(a). This is consistent with the presence of a qubit, frequency tunable between $2.8$~GHz~$<f_\textrm{Q}<4.2$~GHz. The data in Fig.~\ref{fig2}~(b) likely shows the $f_{01}=\fq$, i.e. ground to first excited state transition of the qubit as a function of gate voltage.  

Similar measurements were also performed on device QB and are presented in Fig.~\ref{fig2}~(c) and (d). The resonator of this device exhibits a bare resonance frequency of $f_0=4.595$~GHz for $\Vg>20$~V. When $\Vg$ is tuned below 20~V $f_r$ exhibits upwards shifts, see Fig.~\ref{fig2}~(c). The strongest shift in $f_r$ is observed to be $\sim 7$~MHz. For $\Vg<20$~V a spectroscopic response is visible at the same gate voltages as the shifts in $f_r$ (Fig.~\ref{fig2}~(d)), consistent with the qubit frequency in the range $2.4$~GHz$~<f_\textrm{Q}<3.5$~GHz, cf. Table~\ref{tab:Qubits}.  
\begin{table}[b]
	\centering
	\begin{tabular}{l|c|c|c|c} 
	\hline\hline
 	      & $f_\textrm{0}$ [GHz] & $f_\textrm{Q}$ [GHz] & $g_\textrm{max}$ [MHz] & $\chi_\textrm{max}$ [MHz]  \\ 
	\hline
	Qubit A & 5.572    & $2.8-4.2$  & 113        & 10            \\ 
	\hline
	Qubit B & 4.595    & $2.4-3.5$  & 85         & 7             \\ 
	\hline\hline
	\end{tabular}
	\caption{Parameters extracted from resonator and qubit spectroscopy measurements for devices QA and QB.}
	\label{tab:Qubits}
\end{table}

The coupling strength $g$ between the resonator and the qubit can be estimated using the dispersive shift $\chi$ of the resonator. To first order, due to the coupling to the $f_{01}$ transition, $\chi=\frac{g^2}{\Delta}$, where $\Delta=f_r-\fq$ is the detuning between the qubit and the resonator. Using this expression and the data of QA (Fig.~\ref{fig2}~(a) and (b)) yields an estimation of the coupling strength that shows to increase with observed $\fq$, see Supplementary Information. In particular, we extract a minimum value of $g=48$~MHz and a maximum value of $g=113$~MHz. Similar values are observed with QB, summarized in Table~\ref{tab:Qubits}. 

\begin{figure}[t]
	\centering
	\includegraphics{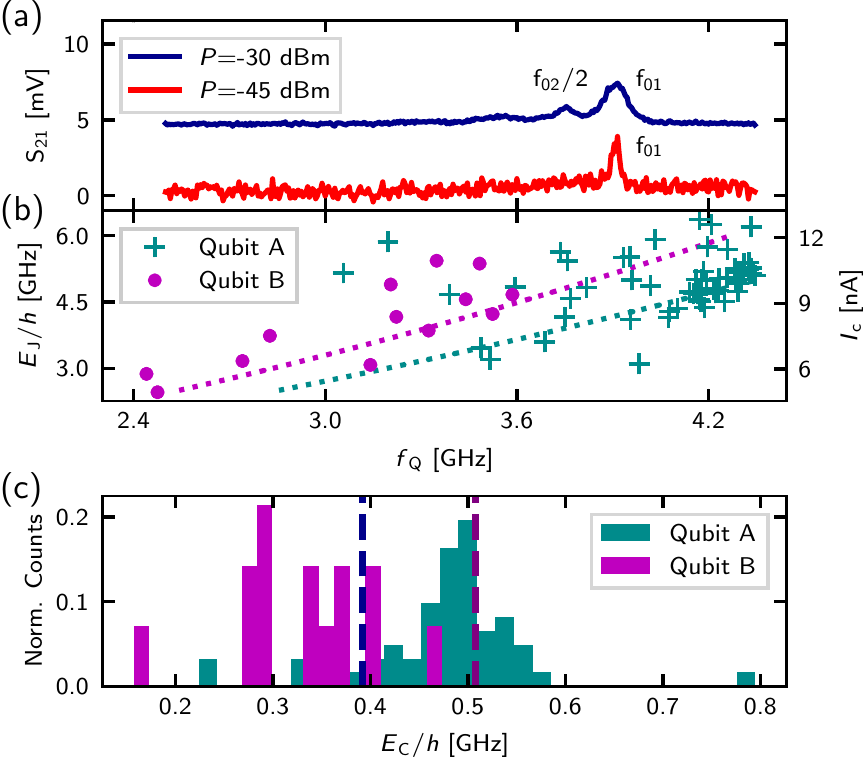}
	\caption{Qubit spectroscopy and qubit parameters $\Ej$ and $\Ec$. (a) Qubit spectroscopy traces at two different qubit drive powers for device \Qthr, offset for clarity. (b) Extracted values for $\Ej$ as a function of $\fq$ for \Qthr~(cyan crosses) and \Qone ~(purple points). The critical current $\Ic$, calculated from $\Ej$ is indicated on the second y-axis. Dotted lines indicate the $\Ej$ values predicted by Eq.~\ref{eq:f01GM}. (c) Normalized histogram of extracted values for $\Ec$. Dashed lines correspond to values of $\Ec$ determined via finite element simulations for the two devices.}
	\label{fig3}
\end{figure}

Mapping out $f_{01}$ is the first step of characterizing the voltage tunable qubit. However, to fully describe the qubit Hamiltonian, $E_\textrm{J}$ and $E_\textrm{C}$ are needed. These parameters cannot be deduced from the qubit's fundamental frequency alone. Here, we show that they can nevertheless be extracted by measuring the qubit at different drive powers. For this purpose, we use the same qubit  spectroscopy technique as in Fig.~\ref{fig2}~(b), holding $\Vg$ fixed and varying the qubit drive power. Such a measurement is presented in Fig.~\ref{fig3}~(a). At a low drive power, $P=-45$~dBm at the output of the microwave generator, only a single peak is observed in the qubit spectroscopy (Fig.~\ref{fig3}~(a) bottom trace). If the power of the qubit drive is increased to $P=-30$~dBm, a more complicated multi-peak response is observed, which exhibits a relatively clear second peak at frequencies just lower than $f_{01}$, see Fig.~\ref{fig3}~(a). This is indicative of a weakly anharmonic circuit, such as a transmon qubit, where the lower peak corresponds to the two photon transition from the ground state to the second excited state, i.e. $f_{02}/2$. Note that this second spectral peak is not always clearly present, and a broader spectral feature is consistently seen at higher drive powers, within which it is not possible to resolve clear individual peaks. This may be due to significant charge dispersion of higher qubit energy levels, or other unknown sources of decoherence.
We measure the frequency of the two clearest spectral lines over a range of $\Vg$ from data similar to that seen in Fig.~\ref{fig3}~(a). Interpreting them as the $f_{01}$ and $f_{02}/2$ transitions of a transmon, we can extract the anharmonicity $\alpha=2\left(f_{01}-f_{02}/2\right)$. By using a first order perturbation theory approach to a modified Cooper pair box Hamiltonian presented by \emph{Kringh\o j et al.}~\cite{Kringhoj2018} $\alpha$ and $f_{01}$ are given by 
\begin{equation}
	h\alpha=\Ec\left(1-\frac{3\Ej}{\Delta N}\right),\\
	\label{eq:anharmGM}
\end{equation}
\begin{equation}
	hf_{01}=\sqrt{8\Ej\Ec}-\Ec\left(1-\frac{3\Ej}{\Delta N}\right),
	\label{eq:f01GM}
\end{equation}
where $\Delta$ is the induced superconducting gap of the CNT JJ (we estimate $\Delta=90~\mu$eV using DC bias spectroscopy measurements on CNT devices contacted with identical processing) and $N$ the number of conduction channels contributing to Cooper-pair transport. Using Eq.~\ref{eq:anharmGM} and Eq.~\ref{eq:f01GM} together with the values extracted for $f_{01}$ and $\alpha$, the qubit parameters $\Ej$ and $\Ec$ can be calculated by solving the resulting set of equations. 

If all channels are assumed to exhibit equal transmission $\mathcal{T}$, Eq.~\ref{eq:anharmGM} can be rewritten using $\Ej=\frac{\Delta}{4}	N\mathcal{T}$ such that $\mathcal{T}=\frac{4}{3}(1-\alpha/\Ec)$~\cite{Kringhoj2018}. The measured $\alpha$ can now be used to calculate $\mathcal{T}$ at each $\Vg$ (or $f_{01}$) resulting in pairs of $\left[\mathcal{T},f_{01}\right]$. Comparing these with $hf_{01}(\mathcal{T})=\sqrt{2\Delta\Ec\mathcal{T}N}-\Ec\left(1-\frac{3}{4}\mathcal{T}\right)$ (resulting from rearranging Eq.~\ref{eq:f01GM}) for $N\in\{1,2,3,4\}$ and $\Ec$ taken from electrostatic modelling for the exact qubit design ($\Ec^{\Qthr}=508$~MHz, $\Ec^{\Qone}=391$~MHz), good agreement was found for $N=1$, see Supplementary Information.

In Fig.~\ref{fig3}~(b) the Josephson energy $\Ej$, extracted with the method mentioned above, is presented as a function of qubit frequency $f_{01}=\fq$. The dashed lines correspond to expected values for $\Ej$ according to Eq.~\ref{eq:f01GM} (with $\Ec$ extracted from simulations). The spread in values is attributed to the error in extracting $f_{02}/2$ from the qubit power spectroscopy data as these can be noisy and sometimes exhibit a complicated multi-peak structure, making the peak distinction difficult. The Josephson energy can also be used to calculate the critical current $\Ic$ of the qubit's JJ using $\Ej=\hbar\Ic/(2e)$, see Fig.~\ref{fig3}~(b). We find $5$~nA~$<\Ic<13$~nA, with an average value $\langle I_\textrm{c}\rangle=9.6\pm 1.6$~nA across both qubit devices. These values of $\Ic$ are comparable to $1$~nA~$<\Ic<16$~nA that we independently observed in DC measurements of CNT JJs. The normalized histogram for the values of $\Ec$, extracted simultaneously with the values of $\Ej$, is presented in Fig.~\ref{fig3}~(c). Dashed vertical lines correspond to simulated values of $\Ec$, as mentioned above, and show good agreement with the data.

The obtained values of $\Ej$ and $\Ec$ can be used to calculate the ratio $\EjEc$. We find a  mean $\EjEc\approx 11$ for \Qthr~and $\EjEc\approx 13$ for \Qone. Again, these values agree well with electrostatic simulations of the device design, yielding $\EjEc\sim 12$, giving confidence in the extraction method of $\Ej$ and $\Ec$. Note that the estimated average $\EjEc$ ratio places the qubit between the transmon regime $(\EjEc>20)$ and the Cooper pair box regime $(\EjEc<1)$, where a low $\EjEc$ leads to a complex energy level structure with charge dispersion making the qubit susceptible to charge noise~\cite{Koch2007}. However, charge dispersion is predicted to vanish in channels where the transmission is approaching unity~\cite{Averin1999}, which is the case for JJs made from CNTs.

We finally report on investigations of the coherence of the CNT-qubit devices. Conclusive measurements of Rabi oscillations were not observed, likely due to poor coherence (see Supplementary Information). We therefore resort to an alternative pulsed technique for measuring the relaxation time, previously used in quantum dot charge qubits~\cite{Petta2004,Penfold-Fitch2017}. The method was tested on standard superconducting qubits to confirm that it yields the same result as standard techniques (see Supplementary Information).

\begin{figure}
	\centering
	\includegraphics{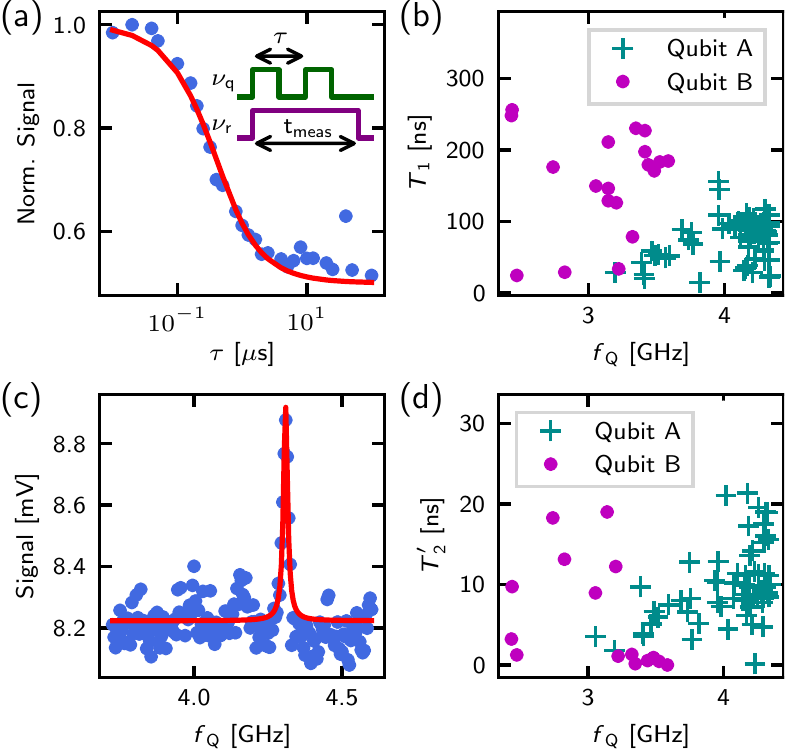}
	\caption{$T_1$ and $T_2^\prime$ measurements. (a) Data of a single $T_1$ experiment on device QA. The measurement response is fitted to Eq.~\ref{eq:T1form} (red curve), yielding $T_1=117.3\pm5.8$~ns. Inset: Pulse scheme (qubit pulse - green, measurement pulse - purple). (b) Measured $T_1$ as a function of $f_\textrm{Q}$ for devices QA (crosses) and QB (circles). (c) A single measurement of $T_2^\prime$ on device QA. Qubit spectroscopy trace is fitted to Eq.~\ref{eq:T2p} (red curve), yielding $T_2^\prime=19.0\pm1.5$~ns. (d) Measured $T_2^\prime$ as a function of $f_\textrm{Q}$ for devices QA (crosses) and QB (circles).}
	\label{fig4}
\end{figure}

To measure $T_1$, a pulse chopping method~\cite{Petta2004,Penfold-Fitch2017} is used. The resonator is continuously measured with a weak cavity drive at $f_0$ for a time of 100~$\mu$s. Simultaneously, the qubit is driven on resonance with a pulse train exhibiting a 50\% duty cycle and for each measurement the pulse period $\tau$ is varied, see inset Fig.~\ref{fig4}~(a). For very short $\tau$, i.e. $\tau\ll T_1$, the qubit drive randomizes the qubit between the ground and first excited state and it has no time to relax. In the case of very long $\tau$, i.e. $\tau\gg T_1$, the qubit has time to relax to the ground state in between drive pulses. Therefore, in the latter case, the measured signal is the time average of the qubit being in the ground and excited state, giving a signal of half the value found in the limit $\tau\rightarrow 0$. A measurement following this procedure is presented in Fig.~\ref{fig4}~(a). The data is normalized with respect to a measurement with the qubit drive turned off and fitted to
\begin{equation}
	S(\tau)=\frac{1}{2}+\frac{T_1(1-e^{-\tau/(2T_1)})}{\tau}
	\label{eq:T1form}
\end{equation}
where $T_1$ is the only free parameter. In the measurement shown in Fig.~\ref{fig3}~(a) the fit yields $T_1=117.3\pm5.8$~ns.
The relaxation time $T_1$ was measured across a range of gate voltages, i.e. $f_\textrm{Q}$, for both devices, see Fig.~\ref{fig4}~(b). QB on average exhibits a longer $\langle T_{1,QB}\rangle=151\pm 71$~ns than QA where $\langle T_{1,QA}\rangle=74\pm 30$~ns. Largest $T_1$ values of 250~ns and 150~ns correspondingly for QA and QB were observed
 
A lower bound, $T_2^\prime$, on qubit coherence time, $T_2$ can be found by measuring the linewidth of a low power qubit spectroscopy trace~\cite{Schuster2005,Abragam1961}, see Fig.~\ref{fig4}~(c). Fitting a Lorentzian with linewidth $2\delta\nu_\textrm{HWHM}$ to the qubit transition peak, $T_2^\prime$ can be calculated via
\begin{equation}
	2\pi\delta\nu_{\textrm{HWHM}}=\frac{1}{T_2^\prime}=\left(\frac{1}{T_2^2}+n_\textrm{s}\omega^2_\textrm{vac}\frac{T_1}{T_2}\right)^{1/2}
	\label{eq:T2p}
\end{equation}
where $n_\textrm{s}\omega^2_\textrm{vac}$ is proportional to the microwave input power~\cite{Schuster2005,Abragam1961}. Hence, at low qubit drive powers the linewidth should be the least broadened. This results in $T_2^\prime=19.0\pm1.5$~ns for the data presented in Fig.~\ref{fig4}~(c). 

The measurement and its analysis was repeated for different qubit frequencies $f_\textrm{Q}$, see Fig.~\ref{fig4}~(d). While $T_2^\prime$ seems to increase slightly with increasing $f_\textrm{Q}$ for QA, this is not true for QB. Coherence is highest at around $f_\textrm{Q}=3$~GHz, with $T_2^\prime=25$~ns for QB, but significantly reduced at $f_\textrm{Q}=3.5$~GHz. On average QA exhibits a longer $\langle T_{2,QA}^{\prime}\rangle=10\pm 5$~ns compared to QB $\langle T_{2,QB}^{\prime}\rangle=6\pm 6$~ns. Although these coherence times seem rather short, these only represent a lower bound for $T_2$. However, the low coherence times could be attributed to dissipation due to dirty, disordered CNTs, Purcell decay into the gate line and residual resistance to the superconducting leads. Dissipation in nanoscale weak link JJ oscillators, made from aluminium, was previously mentioned as a possible source of decoherence~\cite{Vijay2009,Vijay2010,Levenson-Falk2011}. Additionally, the short $T_2^\prime$ could also stem from Andreev levels in the junction interacting with acoustic phonons~\cite{Zazunov2005,Janvier2015a,Gramich2015a}.

The experiments described here demonstrate a first implementation of a voltage tunable superconducting qubit based on a CNT JJ. The device is of similar geometry and exhibits similar gate voltage behaviour to previously reported voltage tunable superconducting qubit devices~\cite{DeLange2015,Larsen2015a,Casparis2018a,Kroll2018a,Wang2018}. Simultaneous resonator and qubit spectroscopy showed clear evidence of qubit-resonator coupling with coupling strength on the order of $g\sim 100$~MHz, comparable to state of the art cQED experiments with conventional transmon qubits. Qubit spectroscopy at high drive powers was used to extract the qubit parameters $\Ej$, $\Ec$, their ratio $\Ej/Ec$ and $\Ic$ of the JJ. Further, qubit relaxation and coherence times in the range $10-200$~ns were observed.

Although the coherence of these first devices is not competitive with state-of-the-art superconducting qubits, advances in fabrication, e.g. using suspended ultra-clean CNT JJs could lead to significant improvements. Such JJs have already been individually realized~\cite{Schneider2012}. Additionally, growing CNTs from purified C$_{12}$ methane would offer a platform for ballistic and defect free JJs, potentially having a huge impact on coherence of the qubits. Fabrication improvements in other hybrid qubit designs, such as those based on InAs nanowires resulted in $T_1=5-20~\mu$s~\cite{Casparis2016,Luthi2018}, similar to state-of-the-art, aluminum-based flux-tunable transmon qubits~\cite{Kelly2015}. The implementation of a superconducting qubit with a CNT presented here offers potential for unique experiments in order to create quantum interference between a qubit and mechanical motion~\cite{Khosla2018}. Additionally, CNT-based qubits could be used as ultrasensitive force sensors~\cite{Moser2013} and if arranged in a SQUID geometry as a detector for magnetic moments~\cite{Cleuziou2006}. Furthermore, these qubits based on proximitized CNTs could be utilized to study Andreev physics~\cite{Gramich2017,Tosi2019,Janvier2015} and investigate the prediction of carrying Majorana fermions~\cite{Klinovaja2012,Sau2013,Marganska2018}, which could be valuable for topological quantum computing. 

\section{Methods}
Device fabrication begins with CNTs grown via chemical vapour deposition on a Si/SiO$_2$ (450~$\mu$m/300~nm) substrate. The microwave resonators are then patterned via electron beam lithography (EBL). Prior to metal deposition an oxygen plasma etch is carried out to remove any CNTs that might electrically short the microwave circuits. Afterwards, 100~nm of Al is deposited via electron beam evaporation. Following lift-off, SEM imaging is used to locate and select CNTs for the qubits. The contacts to the CNT and island of the qubit are then patterned with EBL, post-development cleaned using UV ozone and metalized with a Pd/Al (4/80~nm) bi-layer. Before sample mounting, the room-temperature resistances of the CNT JJs are measured to check the fabrication yield. Roughly 80\% of the fabricated devices conduct at room-temperature and exhibit resistances between 7~k$\Omega<R_n<100$~k$\Omega$. A series of 3 chips were fabricated, each consisting of 10 potential qubits. Each chip contains 10 $\lambda/4$ resonators with different frequencies, multiplexed via capacitive coupling through a single microwave transmission line, see Fig.~\ref{fig1}~(a). All chips are based on the same architecture, except that on the chip of QA the capacitive coupling of the resonator to the transmission line was increased, compared to the chip of QB.

\section{Acknowledgements}
\acknowledgements{We acknowledge support from the Royal Academy of Engineering, EPSRC (EP/R029229/1), Swiss Nanoscience Institute (SNI) and the Swiss National Science Foundation. M. M. acknowledges support from the Stiftung der Deutschen Wirtschaft (sdw).}

\section{Author contributions}
A.B. grew the CNTs at facilities of C.S., A.N. developed the Pd/Al bi-layer and M.M. fabricated the devices. A.P., M.E. assisted in the experimental setup, M.M. performed the measurements and M.M. and P.J.L analyzed the measurements. The Manuscript was prepared by M.M. with P.J.L. and E.A.L, A.N., A.P., M.E. and A.B. providing input. G.A.D.B., E.A.L. and P.J.L supervised the project.


%

\end{document}


\parbox[c]{\textwidth}{\protect \centering \Large \MakeUppercase{Supplementary Information}}
\rule{\textwidth}{1pt}
\vspace{1cm}

\title{Realization of a Carbon-Nanotube-Based Superconducting Qubit}

\author{Matthias Mergenthaler}
\email{mme@zurich.ibm.com}
\affiliation{Clarendon Laboratory, Department of Physics, University of Oxford, Oxford OX1 3PU, United Kingdom}
\affiliation{Department of Materials, University of Oxford, Oxford OX1 3PH, United Kingdom}
\affiliation{IBM Research Zurich, S\"{a}umerstrasse 4, 8803 R\"{u}schlikon, Switzerland}

\author{Ani Nersisyan}
\affiliation{Clarendon Laboratory, Department of Physics, University of Oxford, Oxford OX1 3PU, United Kingdom}
\affiliation{Rigetti Computing, 2919 Seventh Street, Berkeley, CA 94710}

\author{Andrew Patterson}
\affiliation{Clarendon Laboratory, Department of Physics, University of Oxford, Oxford OX1 3PU, United Kingdom}

\author{Martina Esposito}
\affiliation{Clarendon Laboratory, Department of Physics, University of Oxford, Oxford OX1 3PU, United Kingdom}

\author{Andreas Baumgartner}
\affiliation{Department of Physics, University of Basel, Klingelbergstrasse 82, CH-4056 Basel, Switzerland}

\author{Christian Schönenberger}
\affiliation{Department of Physics, University of Basel, Klingelbergstrasse 82, CH-4056 Basel, Switzerland}

\author{G. Andrew D. Briggs}
\affiliation{Department of Materials, University of Oxford, Oxford OX1 3PH, United Kingdom}

\author{Edward A. Laird}
\affiliation{Department of Physics, Lancaster University, Lancaster LA1 4YB, United Kingdom}
\affiliation{Department of Materials, University of Oxford, Oxford OX1 3PH, United Kingdom}

\author{Peter J. Leek}
\email{peter.leek@physics.ox.ac.uk}
\affiliation{Clarendon Laboratory, Department of Physics, University of Oxford, Oxford OX1 3PU, United Kingdom}
	
\date{\today}

	\setcounter{equation}{0}
	\setcounter{figure}{0}
	\setcounter{table}{0}
	\setcounter{page}{1}
	\makeatletter
	\renewcommand{\theequation}{S\arabic{equation}}
	\renewcommand{\thefigure}{S\arabic{figure}}
	\renewcommand{\thetable}{S\Roman{table}}
	\renewcommand{\bibnumfmt}[1]{[S#1]}
	\renewcommand{\citenumfont}[1]{S#1}

\maketitle

\section{Additional Device}

\begin{figure}[hbt]
	\centering
	\includegraphics{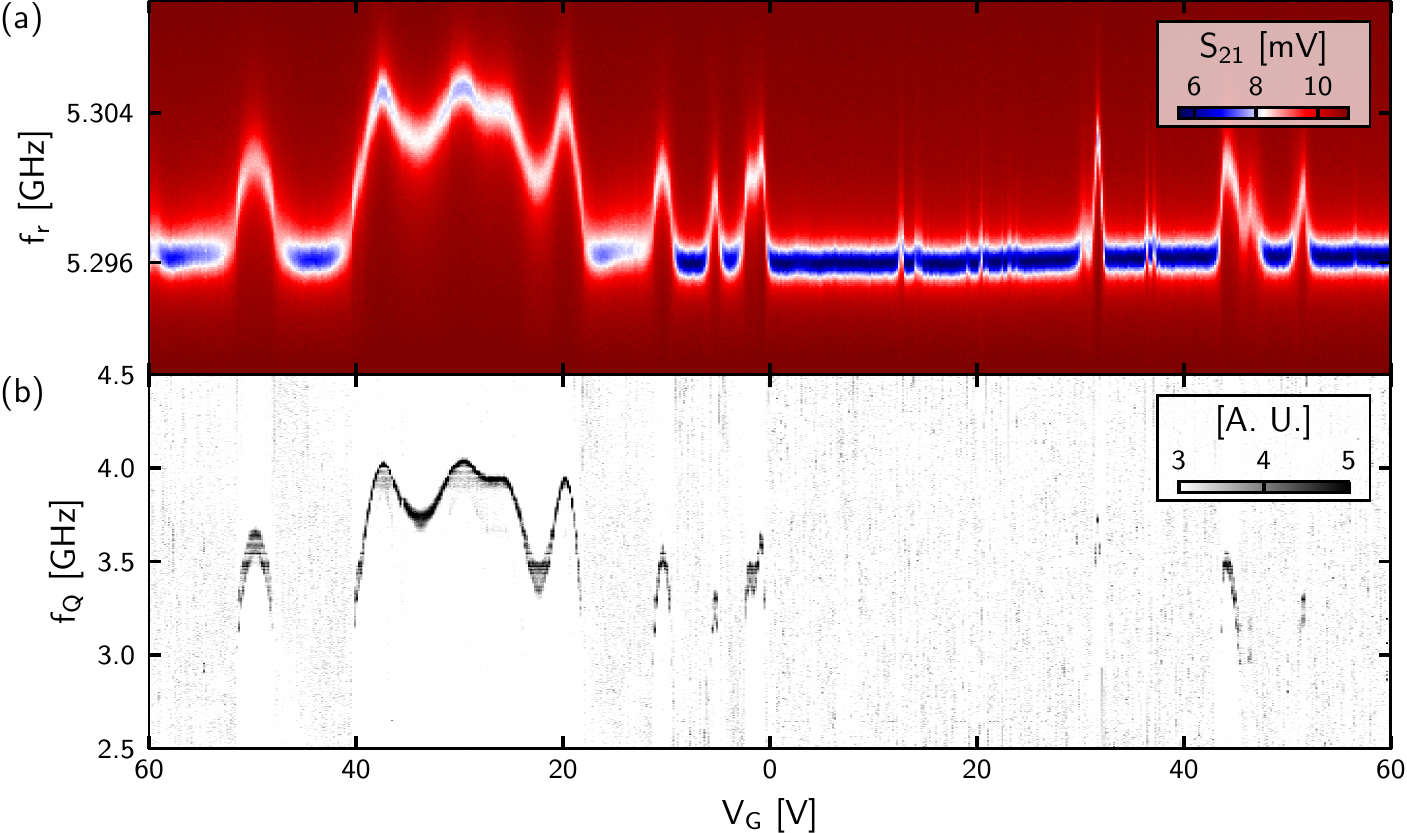}
	\caption{Resonator (a) and qubit (b) spectroscopy as a function of applied gate voltage for a third.}
	\label{ap:fig:respecQ5}
\end{figure}

In this section measurements from an additional qubit device are presented, i.e. resonator and qubit spectroscopy. Figure~\ref{ap:fig:respecQ5} shows the resonator and qubit spectroscopy described in the main text for one more qubit device. The behaviour is similar to the devices presented in the main text. 

Here, the resonator spectroscopy exhibits upwards shifts in the resonance frequency at specific gate voltages. Simultaneously measured qubit spectroscopy shows a clear magnitude response at the same gate voltages, indicating the presence of a qubit with given ground-state $\ket{g}$ to excited-state $\ket{e}$ transition frequency, cf. analysis in the main text.

\section{Coupling Strength}

\begin{figure}[hbt]
	\centering
	\includegraphics{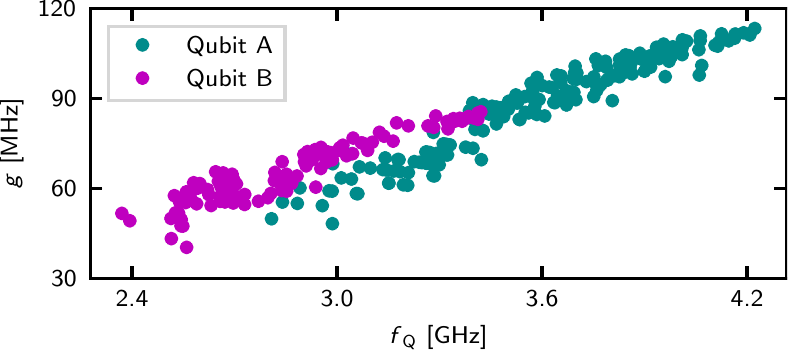}
	\caption{Coupling strength $g$ as a function of qubit frequency $\fq$ for Qubit A and Qubit B.}
	\label{ap:fig:gvsfq}
\end{figure}

Figure~\ref{ap:fig:gvsfq} shows the coupling strength of both qubits, extracted to first order as described in the main text, as a function of qubit frequency $\fq$. An approximately linear dependence on $\fq$ is observed. 

\section{Determining Number of Channels}

\begin{figure}[hb]
	\centering
	\includegraphics{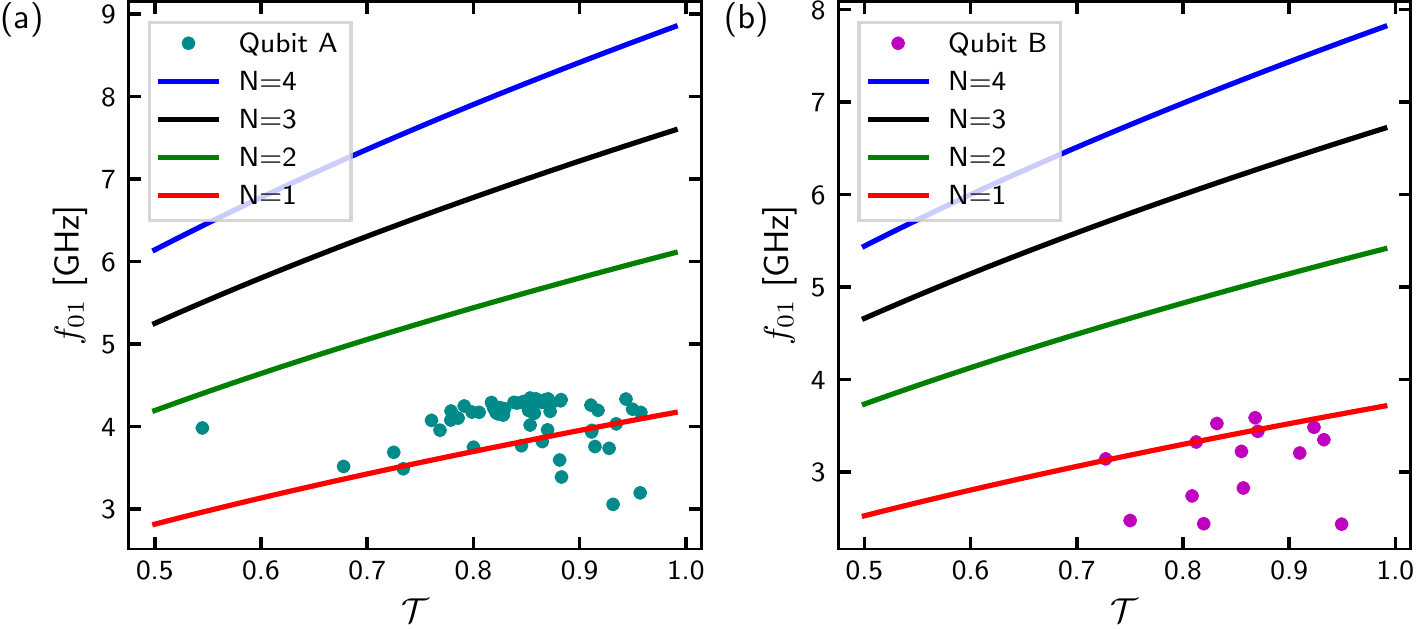}
	\caption{Ground to excited state transition frequency $f_{01}$ as a function of channel transmission $\mathcal{T}$ for Qubit A (a) and Qubit B (b). Curves correspond to calculated values for $N\in\{1,2,3,4\}$ according to Eq.~\ref{sup:eq_f01}.}
	\label{sup:figEcAna}
\end{figure}

When extracting the values for $\Ej$ and $\Ec$ from the data presented in the main text, an assumption has to be made regarding the number of channels $N$ contributing to cooper pair transport. As described in the main text, if all channels are assumed to exhibit equal transmission $\mathcal{T}$, Eq.~1 of the main text can be rearranged using $\Ej=\frac{\Delta}{4}	N\mathcal{T}$ such that~\cite{Kringhoj2018}
\begin{equation}
	\mathcal{T}=\frac{4}{3}(1-\alpha).
\end{equation}
From extracted values of $f_{01}$ and $f_{02}/2$, $\alpha$ can be obtained and used to calculate $\mathcal{T}$ at each $\Vg$ resulting in pairs of $\left[\mathcal{T},f_{01}\right]$. These can be compared with 
\begin{equation}
	f_{01}(\mathcal{T})=\sqrt{2\Delta\Ec\mathcal{T}N}-\Ec\left(1-\frac{3}{4}\mathcal{T}\right),
	\label{sup:eq_f01}
\end{equation}
resulting from rearranging Eq.~2 in the main text, for $N\in\{1,2,3,4\}$ and $\Ec$ taken from electrostatic modelling. This comparison is displayed in Figure~\ref{sup:figEcAna} for both Qubit A as well as Qubit B, showing good agreement only for $N=1$. 

\section{Comparison of Coherence Measurements}

\begin{figure}[h!t]
	\centering
	\includegraphics{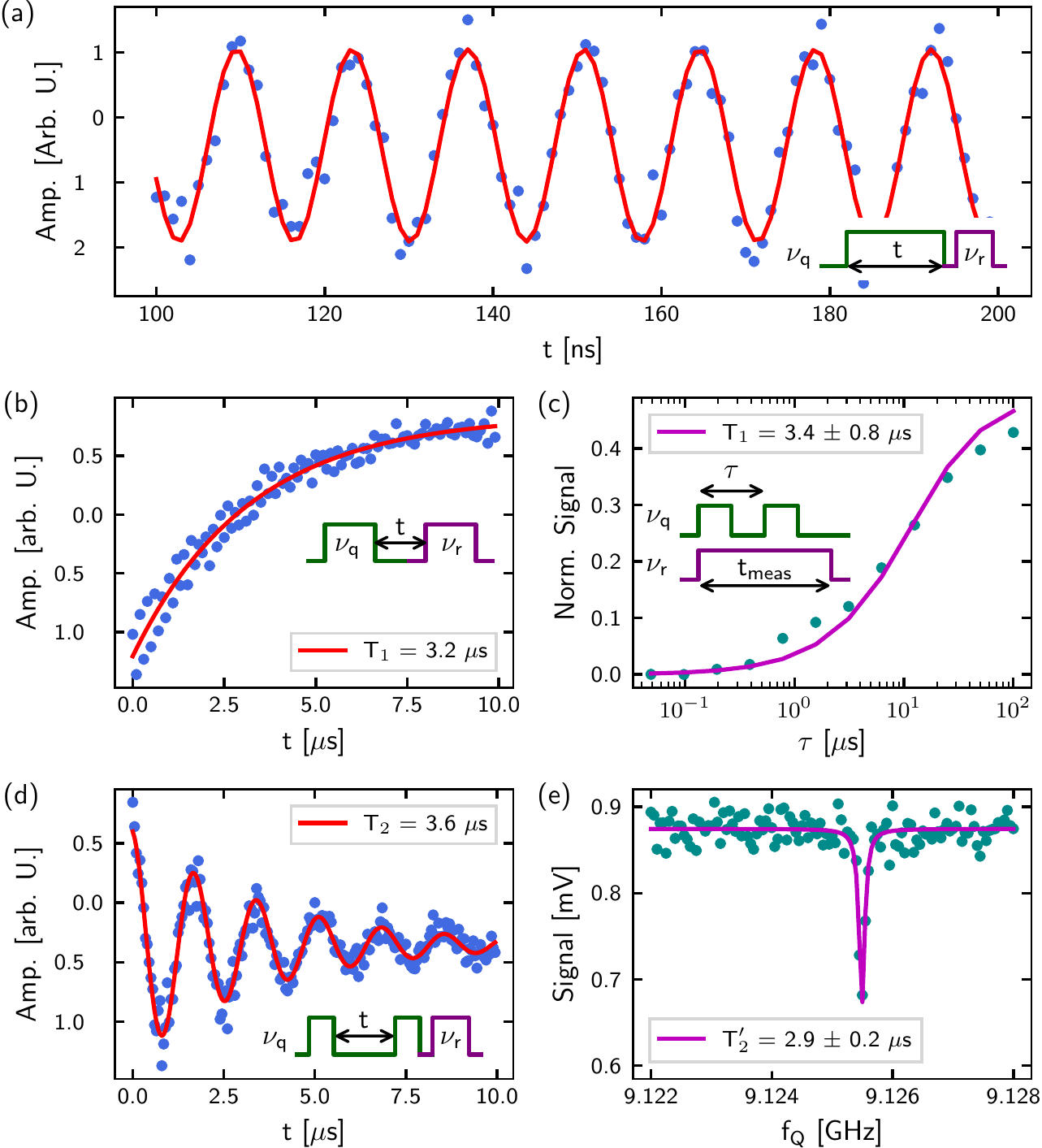}
	\caption{Comparison of $T_1$ and $T_2$ measurement techniques for a transmon qubit in a 3D-cavity. (a) Rabi oscillations of the transmon qubit with sinusoidal fit (red curve). Inset: Pulse scheme to measure Rabi oscillations (qubit pulse - green, measurement pulse - purple). (b) Measurement of $T_1$ with a standard $\pi$-pulsed measurement. Red curve is an exponential decay fit, yielding $T_1=3.2~\mu$s. Inset: Pulse scheme to measure $T_1$ (qubit pulse - green, measurement pulse - purple). (c) Measurement of $T_1$ via chopped qubit microwave pulses with fit (purple curve) resulting in $T_1=3.4\pm0.8~\mu$s. Inset: Pulse scheme to measure $T_1$ with chopped microwave pulses whilst continuously measuring the resonator (qubit pulse - green, measurement pulse - purple). (d) Measurement of $T_2$ via standard Ramsey fringes, with fit to an exponentially decaying sine resulting in $T_2=3.6~\mu$s. Inset: Pulse scheme to measure Ramsey fringes (qubit pulse - green, measurement pulse - purple). (e) Estimating $T_2$ with a measurement of $T_2^\prime$. A Lorentzian (purple curve) is fit to the low power qubit spectroscopy and its linewidth extracted, resulting in $T^\prime_2 = 2.9 \pm 0.2~\mu$s.}
	\label{fig:ap-3DCoh}
\end{figure}

Here a transmon qubit in a 3D-cavity is used in order to compare the coherence measurement techniques presented in the main text to standard time-domain measurements utilising $\pi-$ and $\pi/2-$pulses, see Figure~\ref{fig:ap-3DCoh}. For the latter, a qubit pulse of varying duration $t$ is sent to the cavity at a fixed power and immediately after a measurement pulse is sent, in order to read the state of the qubit. This results in Rabi oscillations, see Figure~\ref{fig:ap-3DCoh}~(a). These oscillations are then used to determine the pulse length of a $\pi$- and $\pi/2$-pulse, needed for the further analysis. In order to now determine $T_1$ of the qubit, a $\pi$-pulse is sent to the cavity and after the wait time $t$ the state of the qubit is measured. Varying the wait time $t$ results in an exponential decay of the qubit's excited state, see Figure~\ref{fig:ap-3DCoh}~(b). This data can be fit with an exponential decay resulting in $T_1=3.2$~$\mu$s. To measure the decoherence time $T_2$ of the qubit, two $\pi/2$-pulses separated by a time $t$ are sent to the qubit and immediately after the second pulse the qubit state is measured (Ramsey interferometry). If the qubit pulse is slightly ($\sim 5$~MHz) detuned from the qubit frequency and the pulse delay $t$ is varied, this results in exponentially decaying Ramsey fringes, see Figure~\ref{fig:ap-3DCoh}~(d). These Ramsey fringes can be fit with an exponentially decaying sine function, resulting in a $T_2=3.2~\mu$s.

The above described method is the standard way of characterising a qubit's lifetime $T_1$ and coherence time $T_2$. However, in certain scenarios it is not possible to tune up a $\pi$-pulse through Rabi oscillations, which might either be due to very low coherence of the qubit or due to a mismatch between the internal and external quality factor of the measurement resonator. For the carbon nanotube superconducting qubits described in the main text, both of these could be the case. Hence, a different measurement technique for measuring $T_1$ and estimating $T_2$ has been used. Here, it is shown that both of these techniques lead to the same result. 

In order to measure $T_1$ a method is borrowed from the field of quantum dot charge qubits, where microwave pulses at the qubit frequency are chopped up into a 50\% duty cycle while the resonator is continuously measured~\cite{Petta2004,Penfold-Fitch2017}. The period $\tau$ of the qubit pulse sequence is increased until only one period fits into the measurement cycle. Care was taken to filter out values of $\tau$ which are integer multiples of the downconversion intermediate frequency, $f_\textrm{IF}=125$~MHz, as leakage through an IQ mixer can lead to false data points at these pulse periods. The longer the qubit pulse period $\tau$ the more time the qubit has to relax. The resulting decay is normalised to the signal without a qubit drive on and fit to
\begin{equation}
	S(\tau)=\frac{1}{2}-\frac{T_1(1-e^{-\tau/(2T_1)})}{\tau},
\end{equation}\label{eq:ap-T1chop}
see Figure~\ref{fig:ap-3DCoh}~(c)\footnote{It is worth mentioning that Eq.~\ref{eq:ap-T1chop} has a minus sign instead of a plus sign, cf. Eq.~3 in the main text, as the measurement is done in reflection rather than in transmission.}. This results in $T_1=3.4\pm0.8~\mu$s, where the value extracted from the standard pulsed measurement is within the error. The inaccuracy in the fit and the large error is due to the qubit's $T_1$ being quite long and the signal not being able to tail off completely due to restrictions in the measurement acquisition time.

The coherence time $T_2$ can be estimated from the low power linewidth of the qubit spectroscopy~\cite{Schuster2005}. For this the linewidth is fit with a Lorentzian for different qubit drive powers. For low enough powers the linewidth should not change and $1/T_2^\prime=2\pi\delta\nu_{\textrm{HWHM}}=(1/T_2^2+n_\textrm{s}\omega^2_\textrm{vac}T_1/T_2)^{1/2}$~\cite{Schuster2005,Abragam1961}. This indicates that $T_2^\prime<T_2$ and that $T_2^\prime$ is a lower bound for $T_2$. By fitting the lowest power qubit spectroscopy with a Lorentzian, see Figure~\ref{fig:ap-3DCoh}~(e), we get $T_2^\prime=2.9\pm0.2~\mu$s, which is fairly close to the value measured with Ramsey fringes and gives a reasonable lower bound for $T_2$. 

Given the above results, we can conclude that the measurement techniques used in the main text in order to estimate the qubit's lifetime and coherence time are valid alternatives over the standard pulsed approach.

\section{Signatures of Rabi Oscillations}

Although it was not possible in the presented qubit devices to measure Rabi oscillations with separated qubit and measurement pulses, visibility of Rabi oscillations has been previously reported with weak continuous measurements while applying the qubit drive tone~\cite{Wallraff2005,Vijay2012}. A similar type of measurement was performed and some potential evidence of Rabi oscillations observed. 

\begin{figure}[h!t]
	\centering
	\includegraphics{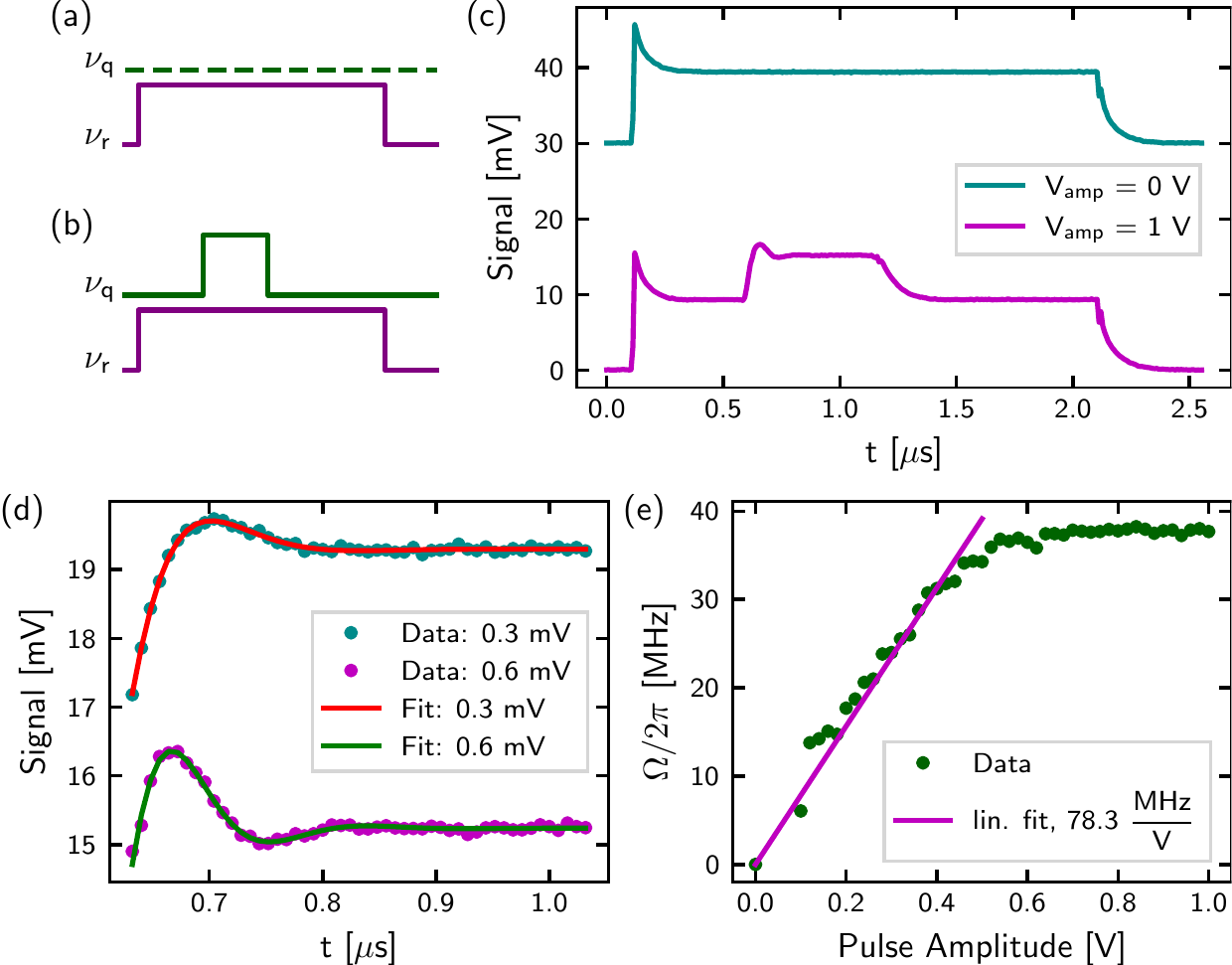}
	\caption{Signatures of Rabi oscillations in device v3\_Q3. (a) Pulse scheme with a $2~\mu$s long resonator pulse (purple) and no qubit drive pulse. (b) The same pulse scheme, now with a 500~ns qubit drive pulse applied during the middle of the measurement pulse. (c) Time-domain resonator response for pulse scheme in (a) top curve (cyan, offset by 30~mV for clarity) and pulse scheme in (b) bottom curve (purple). (d) Zoom-in of the response of the resonator due to the qubit pulse for two different qubit drive amplitudes (offset for clarity). The two responses exhibit oscillations at different frequencies, both fit to an exponentially decaying sine function (solid lines). (e) Oscillation frequency $\Omega/2\pi$ as a function of the qubit pulse amplitude. For a pulse amplitude $<0.6$~V, $\Omega/2\pi$ is fit with a linear function (solid purple line) yielding a slope of $78.3$~MHz/V.}
	\label{fig:rabi_1}
\end{figure}

The measurements presented in the following paragraphs further investigated the time-resolved response of the resonator, see Figure~\ref{fig:rabi_1}. Specifically, the difference between two cases was considered. One where a measurement pulse is sent to the resonator without a qubit drive pulse (Figure~\ref{fig:rabi_1}~(a)) and another one where during the pulse sent to the resonator a qubit pulse is switched on for 500~ns (Figure~\ref{fig:rabi_1}~(b)). In the second case it was possible to vary the amplitude V$_\textrm{amp}$ of the qubit pulse. In both cases the power of the measurement pulse was kept as low as possible in order to prevent the resonator pulse from disturbing the qubit. The measured time responses of the resonator for both pulse schemes are presented in Figure~\ref{fig:rabi_1}~(c). The top curve corresponds to the pulse scheme with no qubit drive pulse, i.e. V$_\textrm{amp}=0$~V. Here a measurement tone is turned on for 2~$\mu$s and the turn on of the resonator is initially observed until it relaxes in its steady state population. When the pulse is turned off the photon population of the resonator decays with its decay rate $\kappa$. The bottom curve of Figure~\ref{fig:rabi_1}~(c) corresponds to the pulse scheme with a $500$~ns long qubit drive tone applied during the measurement pulse. The response initially exhibits identical behaviour compared to the resonator response without a qubit drive. However, when the qubit drive is turned on at $t=0.5~\mu$s the resonator exhibits a stronger magnitude response as the qubit is excited. The response due to the qubit drive pulse seems to display one period of initial oscillation before it reaches the steady state. When the qubit pulse is turned off, the resonator response due to the qubit pulse decays on a timescale comparable to $T_1$. Eventually, once the resonator pulse is turned off, the resonator population again decays at a rate $\kappa$ as for the case without a qubit pulse.

The amplitude of the qubit drive pulse can be varied from 0~V$~<V_\textrm{amp}<1$~V. Further examining the resonator response due to the qubit pulse, it is obvious that the initial oscillation is only present for the first 150-200~ns after the qubit drive pulse is turned on, see Figure~\ref{fig:rabi_1}~(d). Additionally, varying the qubit pulse amplitude results in a modulation of the oscillation frequencies of this initial response to the qubit drive pulse. Figure~\ref{fig:rabi_1}~(d) shows a close up of the response due to the qubit drive pulse for two different amplitudes. The top curve, i.e. for a lower qubit drive pulse amplitude, exhibits a slower oscillation compared to the bottom curve for double the qubit pulse amplitude. Both of these measurements are fit with an exponentially decaying sine curve, each yielding a frequency $\Omega/2\pi$ of the respective oscillation. This analysis can be done for a sweep of the qubit drive pulse amplitude $V_\textrm{amp}$. The extracted $\Omega/2\pi$ is plotted against the qubit pulse amplitude in Figure~\ref{fig:rabi_1}~(e). For pulse amplitudes $<0.6$~V, $\Omega/2\pi$ linearly increases with a slope of $\sim 78.3$~MHz/V, determined from a linear fit (solid purple line). This linear dependence on qubit drive amplitude is a characteristic feature of Rabi oscillations. For pulse amplitudes $>0.6$~V, $\Omega/2\pi$ saturates at just below $40$~MHz which may be explained as being due to the low anharmonicity of the qubit (the qubit will be driven into higher energy states at drive rates greater or equal to the anharmonicity).


%